\documentclass[runningheads,a4paper]{llncs}

\usepackage[USenglish]{babel}
\usepackage[autostyle]{csquotes}

\usepackage{amssymb,amsmath}

\usepackage{nth}
\usepackage{amsfonts}

\usepackage{cite}

\usepackage{tikz}
\newlength{\RoundedBoxWidth}
\newsavebox{\GrayRoundedBox}
   {\setlength{\RoundedBoxWidth}{\dimexpr#1}
    \begin{lrbox}{\GrayRoundedBox}
       \begin{minipage}{\RoundedBoxWidth}}%
   {   \end{minipage}
    \end{lrbox}
    \begin{center}
    \begin{tikzpicture}%
       \draw node[draw=black,fill=black!10,rounded corners,%
             inner sep=2ex,text width=\RoundedBoxWidth]%
             {\usebox{\GrayRoundedBox}};
    \end{tikzpicture}
    \end{center}}

\usepackage{xcolor}
\usepackage{color, colortbl}
\definecolor{Gray}{gray}{0.9}

\usepackage{tikz}
\usepackage{multirow}
\usepackage{hhline}
\usepackage{booktabs}

\usepackage{listings}
\usepackage{fancyvrb}
\usepackage{enumitem}

\usepackage{graphicx}
\usepackage{float}
\usepackage{caption} 
\usepackage{subcaption}
\usepackage{subfig}
\usepackage{placeins}

\newfloat{parallelFigure}{t}{lom}
\restylefloat*{parallelFigure}
\floatstyle{plain}
\floatname{parallelFigure}{Fig.}
\newsubfloat[position=top]{parallelFigure}

\usepackage{url}
\urldef{\mailsa}\path|{lars.ackermann, stefan.schoenig, stefan.jablonski}@uni-bayreuth.de |

\newcommand{\keywords}[1]{\par\addvspace\baselineskip\noindent\keywordname\enspace\ignorespaces#1}

\newcommand{\code}[1]{$#1$}

\title{Inter-Paradigm Translation of Process Models using Simulation and Mining}
\titlerunning{Inter-Paradigm Translation of Process Models using Simulation and Mining}
\author{Lars Ackermann\inst{1}, Stefan Sch{\"o}nig\inst{2} \and Stefan Jablonski\inst{1}}
\authorrunning{Ackermann et al.}
\institute{University of Bayreuth, Germany\\
\email{\{firstname.surname\}@uni-bayreuth.de}\\
\and
Vienna University of Economics and Business,  Austria
}

\begin{document}
\maketitle

\begin{abstract}
Process modeling is usually done using imperative modeling languages like BPMN or EPCs. In order to cope with the complexity of human-centric and flexible business processes several declarative process modeling languages (DPMLs) have been developed during the last years. DPMLs allow for the specification of constraints that restrict execution flows. They differ widely in terms of their level of expressiveness and tool support. Furthermore, research has shown that the understandability of declarative process models is rather low. Since there are applications for both classes of process modeling languages, there arises a need for an automatic translation of process models from one language into another. Our approach is based upon well-established methodologies in process management for process model simulation and process mining without requiring the specification of model transformation rules. In this paper, we present the technique in principle and evaluate it by transforming process models between two exemplary process modeling languages.
 
\end{abstract}

\section{Introduction}
Two different types of processes can be distinguished \cite{Jablonski1994}: well-structured routine processes with exactly predescribed control flow and flexible processes whose control flow evolves at run time without being fully predefined a priori. In a similar way, two different representational paradigms can be distinguished: imperative process models like BPMN\footnote{The BPMN 2.0 standard is available at \url{http://www.omg.org/spec/BPMN/2.0/}.} models describe which activities can be executed next and declarative models define execution constraints that the process has to satisfy. The more constraints we add to the model, the less eligible execution alternatives remain. As flexible processes may not be completely known a priori, they can often be captured more easily using a declarative rather than an imperative modelling approach \cite{VanderAalst2009,Pichler2012,Vaculin2011}. Due to the rapidly increasing interest several declarative languages like \emph{Declare} \cite{pesic2006declarative}, \emph{Dynamic Condition Response} (DCR) Graphs \cite{hildebrandt2013contracts} or \emph{Declarative Process Intermediate Language (DPIL)}  \cite{zeising2014towards} have been developed in parallel and can be used to represent these models. Consequently, flexible processes in organizations are frequently modeled in several different notations.
\begin{parallelFigure}
\centering
\includegraphics[width=\textwidth]{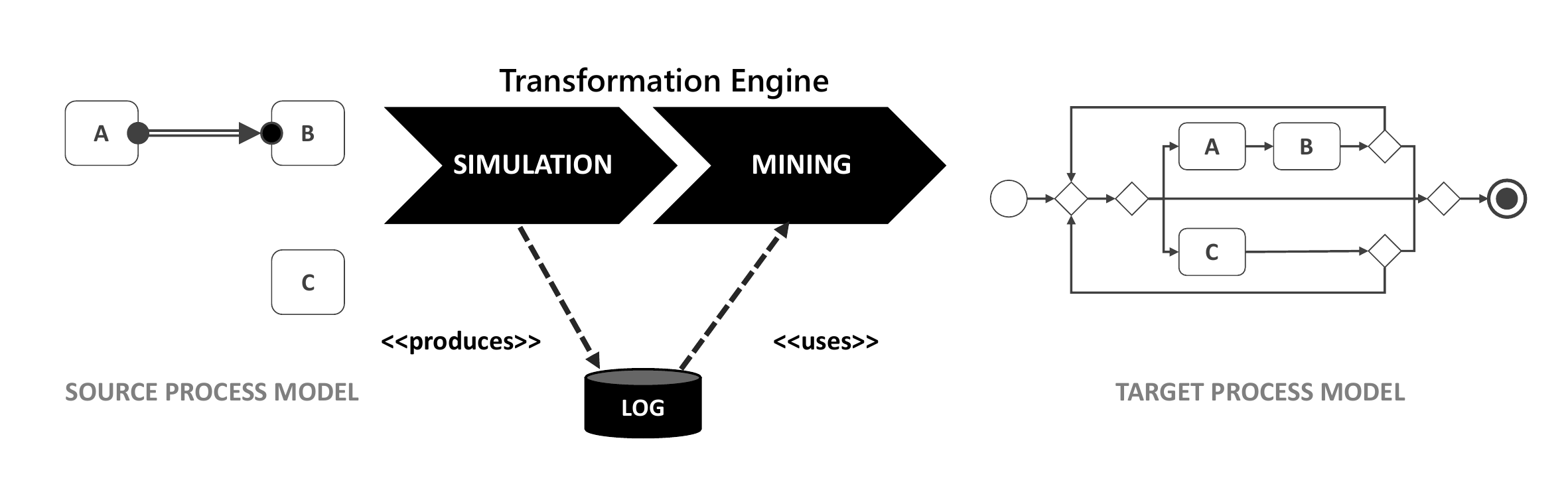}
\caption{Overview of the model transformation approach}
\label{fig:overview}
\end{parallelFigure}
Due to several reasons in many cases a translation of process models to a different language is desired: \textit{(i)} since declarative languages are difficult to learn and understand \cite{Pichler2012}, users and analysts prefer the representation of a process in an imperative notation, \textit{(ii)} even if the user is familiar with a particular notation neither imperative nor declarative languages are superior for all use cases \cite{Prescher2014}, \textit{(iii)} adopted process execution systems as well as analysis tools are tailored to a specific language and \textit{(iv)} since process modeling is an iterative task, the most appropriate representation for the evolving process model may switch from a declarative to an imperative nature and vice versa. To facilitate these scenarios, a cross-paradigm process model transformation technique is needed. While contemporary research mainly focuses on transforming process models between different imperative modeling languages, approaches that comprise declarative languages are still rare~\cite{Prescher2014}.

We fill this research gap by introducing a two-phase, bi-directional process model transformation approach that is based upon existing process simulation and mining techniques. The principle is highlighted in Fig.~\ref{fig:overview}. Model-to-model transformation techniques usually involve the creation of transformation rules which is an error-prone and cumbersome task~\cite{Wimmer2007}. Hence, we created an approach that does not require the definition of transformation rules. First, a set of valid execution traces of the process is automatically generated by simulating the source model. Second, the resulting event log is analyzed with a process mining approach that uses the target language to represent the discovered model. For the work at hand we use Declare as a representative of declarative languages and BPMN as the imperative language. However, note that the approach works with every language framework that provides model simulation and mining functionality. We evaluate functionality and performance by transforming four simplistic examples as well as two real-life process models between BPMN and Declare.

The remainder of this paper is structured as follows: Section \ref{sec:background} describes the fundamentals of declarative process modeling at the example of Declare as well declarative and imperative simulation and mining. In Section \ref{sec:approach} we introduce our approach to transform declarative process models. The approach is evaluated in Section \ref{sec:evaluation}. We discuss related work in Sec.~\ref{sec:relatedwork} and Section \ref{sec:conclusion} concludes the paper.
\section{Background and Preliminaries}
\label{sec:background}
In this section we introduce declarative process modeling as well as the simulation and mining of declarative process models.
\subsection{Declarative Process Modeling}
\label{subsec:DPM}
Research has shown that DPMLs are able to cope with a high degree of flexibility~\cite{Fahland2009}. The basic idea is that, without modeling anything, everything is allowed. To restrict this maximum flexibility, DPMLs like Declare allow for formulating rules, i.e., constraints which form a forbidden region. An example is given with the single constraint \code{ChainSuccession(A,B)} in Fig.\ref{fig:overview}, which means that task B must be executed directly after performing task A. Task C can be performed anytime. The corresponding BPMN model mainly consists of a combination of exclusive gateways. Declare focuses almost completely on control flow and, thus equivalent BPMN models may only consist of control flow elements as well. A brief discussion of issues related to differences in the expressiveness of the two languages is given in Sec.~\ref{subsec:LangLogExpressivness}. Declarative and imperative models are in a manner opposed. If one adds an additional constraint to a declarative model, this usually results in removing elements in the imperative model and vice versa. If, for instance, we add the two constraints \code{Existence(A)} and \code{Existence(C)} to the source process model in Fig.~\ref{fig:overview}, the edge directly leading to the process termination event must be removed. For a transformation approach this means that the identification of appropriate transformation rules would be even more complicated, because a control-flow element in the source language does not necessarily relate to the same set of control-flow elements in the target language in all cases.

\subsection{Process Simulation and Process Mining}
\label{subsec:bpSim}
In this section, we briefly describe the two methods our transformation approach is based on.
Simulating process models is well-known as a cost-reducing alternative to analyzing real-world behavior and properties of business processes~\cite{VanderAalst2010}. Though, there are different simulation types, for our purpose, we exclusively refer to the principle of \textit{Discrete-event Simulation (DES)}~\cite{Stewart2004}. DES is based upon the assumption that all relevant system state changes can be expressed using discrete sequences of events. By implication this means that there is no invisible state change between two immediately consecutive events. This assumption is valid since we use a simulation technique for the purpose of model translation. This means that, in our case, a source process model fully describes the universe of system state changes. For the application in our approach we use simulation techniques to generate exemplary snapshots of process executions allowed by an underlying process model. The produced simulation results are the already mentioned event logs, containing sets of exemplary process execution traces. These logs are then consumed by process mining techniques.

\textit{Process Mining} aims at discovering processes by extracting knowledge from event logs, e.g., by generating a process model reflecting the behaviour recorded in the logs \cite{VanderAalst2011}. There are plenty of process mining algorithms available that focus on discovering imperative process models, e.g., the simplistic \textit{Alpha miner}~\cite{VanderAalst2011} or the \textit{Heuristics Miner}~\cite{weijters2011flexible}. Recently, tools to discover declarative process models like DeclareMiner \cite{Maggi2011} or MINERful~\cite{CiccioM15} have been developed as well. In the approach at hand, we use process mining techniques to automatically model the simulated behaviour in the chosen target language.
\section{Challenges and Preconditions}
\label{sec:challenges}
Our approach at hand requires some prior analysis and raises some challenges we have to deal with. Probably the most important as well as the most trivial challenge is to prevent the transformation approach from causing information loss \textit{(CP1)}. This means that source and target model must be semantically equivalent. However, the approach at hand does not consider interpolation methods but instead assumes that source and target language have the same semantic expressiveness. We therefore provide a limited comparative analysis of the expressiveness of Declare and BPMN in section \ref{subsec:LangLogExpressivness}. \textit{(CP2)} complements the issue of expressiveness. It must be examined whether a process log is expressive enough to be able to cover the behavioral semantics of a process model. While \textit{(CP2)} discusses the general \textit{ability} of log data to preserve the behavioral semantics of a process model, we now have to make sure that it \textit{actually contains} the required execution traces~\cite{weijters2011flexible}. Therefor both transformation steps, simulation as well as process mining, require appropriate parameterizations \textit{(CP3)}. Many process mining approaches suggest that the best parametrization is data-dependent and can therefore be determined in particular only. Hence, it is necessary to provide a strategy for the determination of well-fitting parameter values. 
\section{Contribution} \label{sec:approach}
The translation of a model specified in one language to another is usually done using \textit{mapping rules}. In general, we have a set of $n$ modeling languages that are able to describe the same domain. A translation system that uses this \textit{direct}, mapping-rule-based translation principle has a complexity of $O\left(n\left(n-1\right)\right) = O\left(n^2\right)$ in terms of required rule sets $n$. Finding all rule sets for a system of modeling languages is, therefore, a time-consuming and cumbersome task \cite{Wimmer2007}. On the contrary, our transformation approach is based on the following two core techniques: (i) Process Models Simulation and (ii) Process Mining. Therefore, our approach does not require the creation of transformation rules but uses the core idea to extract the \textit{meaning} of a particular model by generating and analyzing valid instances of the model through simulation. Each simulation result is represented as an event log which is the usual input for process mining techniques such as \cite{weijters2011flexible,Maggi2011,schoenig2015bpmds}. This means that our transformation approach is based on the assumption that we are able to find appropriate simulation and mining technologies for the individual language pairs. In the case of our continuously used BPMN-Declare language pair several simulation and mining techniques are ready to use. The following subsections briefly describe the principles and configuration strategy of these two core techniques. Furthermore, we briefly discuss issues related to the expressiveness of the particular modeling languages and of event log used as a transfer medium.

\subsection{Language and Log Expressiveness}
\label{subsec:LangLogExpressivness}
\FloatBarrier
We have to discuss two key factors for our translation approach: \textit{(i)} Differences in the expressiveness of the particular source and target language and \textit{(ii)} potentially insufficient expressiveness of event logs. 
\begin{parallelFigure}
	\centering
	\includegraphics[width=0.8\textwidth]{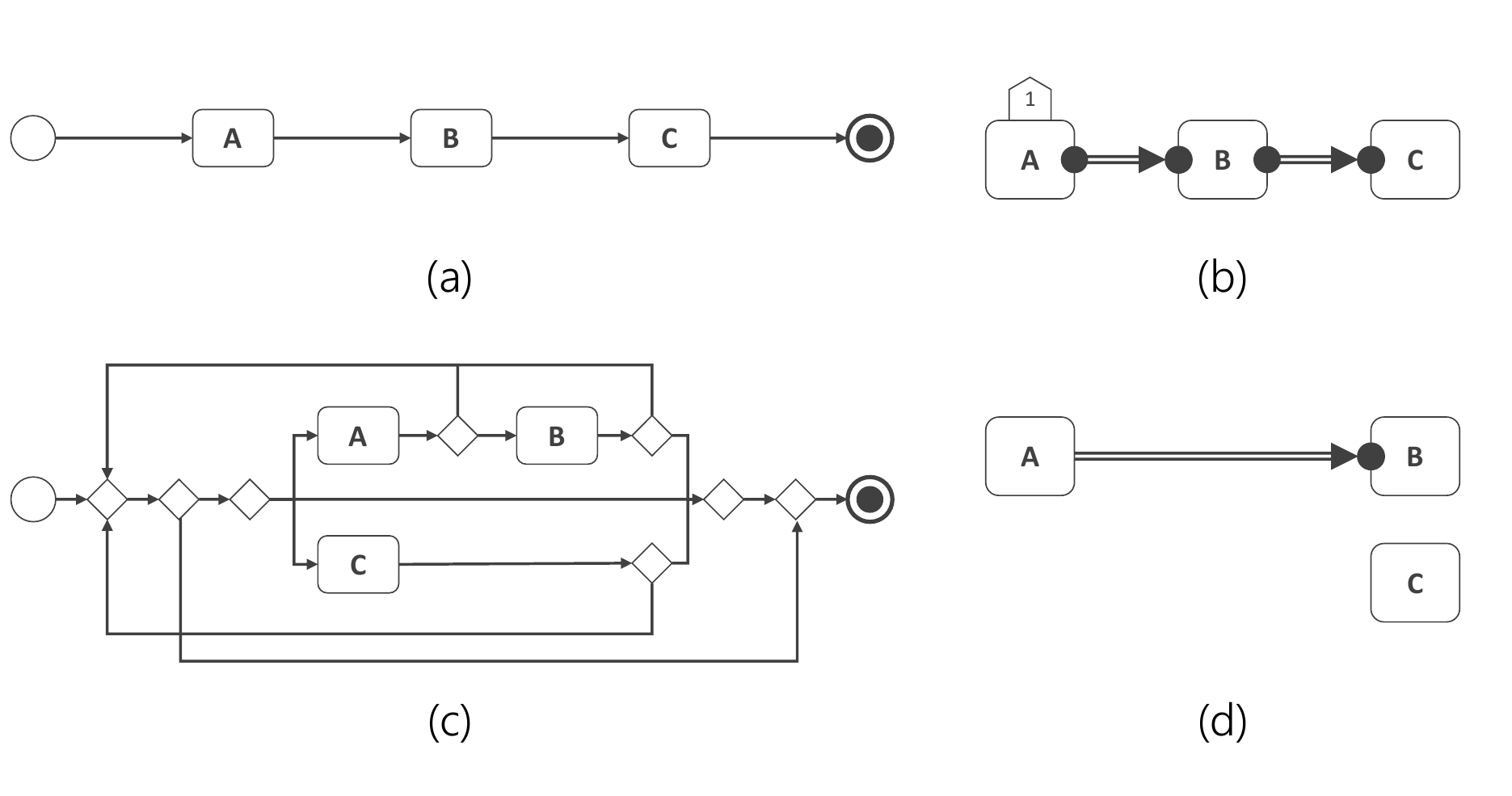}
	\caption{Continuous example}
	\label{fig:contExample}
	\vspace{-15px}
\end{parallelFigure}

Equal \textit{Language Expressiveness} (\textit{CP1}) means, in our context, that two languages, e.g. BPMN and Declare, are able to model the same semantics, no matter if the resulting model is imperative or declarative. Considering our two exemplary process modeling languages, we can easily find significant differences. Even though Declare is extendable, it's expressiveness is currently limited to concepts that describe tasks and temporal or existence constraints. In contrast, BPMN allows for modeling of organizational associations as well as data flow and other elements. In order to provide a profound catalog that describes model patterns which can be translated successfully, an extensive comparison of the two particular process modeling languages is required. Due to the fact that such a deep analysis is currently not available and because this topic would go beyond the scope of this paper we choose example processes for our evaluation that can definitely be represented in both languages. 

The second issue is the question of \textit{Sufficient Log Expressiveness} (\textit{CP2}). An event log \hyphenquote{USenglish}{contains historical information about \hyphenquote{USenglish}{When, How, and by Whome?}}\cite{VanderAalst2015}. An event log describes an example of a process execution and, hence, one possible trace through the source process model. Process mining techniques are built based upon the following assumptions regarding the log contents and structure: \textit{(i)} a process consists of cases that are represented by traces, \textit{(ii)} traces consist of events and \textit{(iii)} events can have attributes like the activity name, a timestamp, associated resources and a transaction type~\cite{VanderAalst2011}. An event can, therefore, unequivocally be associated with the corresponding activity, resources and the type of the system state change. All of these information describe a single state change but not dependencies between state changes. Thus, process mining techniques are limited to the information that can be encoded in sequential, discrete event logs. However, let us consider model \textit{(d)} shown in Fig.~\ref{fig:contExample}. In order to extract this \code{chainPrecedence(A,B)} rule from a process event log, the following condition must be valid for all traces: Whenever an event occurs that refers to activity \code{B} then the event occurring immediately before\footnote{Declare does not distinguish between different transaction types.} must refer to the activity \code{A}. This suggests that temporal relationships can be extracted from the log if the latter's quality and length is sufficient. However, the activity labeled with \code{C} in the same model is not restricted by any constraint. This means, by implication, that it can be executed arbitrarily often. Because a log has a finite length, we cannot encode this knowledge. Instead the mining technique could use some threshold following the assumption, that, if a particular
task has been executed \code{n} times, the number of possible executions is theoretically unlimited. 

 
Like in the case of language expressiveness, a much deeper dive into the limitations of information encoding in discrete-event logs is required but would go beyond the scope of this paper. So far we briefly discussed, what information an event log is \textit{able} to provide. The following three subsections focus on \textit{if} and \textit{how} we can make sure that the desired information are contained in the log. 

\subsection{General Simulation Parameters}
\label{subsec:simParams}
There are two general properties which influence the transformation quality as well as the performance of the whole approach, i.e. \textit{(i) the Number of Traces} $(N)$ and \textit{(ii) the Maximum Trace Length} $(L)$.

Setting the value for $N$ appropriately is the basis for including all relevant paths in the log. Considering the example BPMN model in Fig.~\ref{fig:contExample} (c), there are several gateways whereby each unique decision leads to a new unique execution trace. Hence, we need a strategy for determining the minimum number of traces to include in the log. However, this number depends on the second parameter $L$. Without providing a value for $L$ the simulation of a process model that allows for loops could hypothetically produce traces of infinite length. Thus, the potential number of different traces is also infinite. We therefore need an upper bound for $L$. The lower bound is governed by the process model itself.

The appropriate setting (\textit{CP3}) of the introduced parameters depend on the source process model. In the case of the BPMN model in \ref{fig:contExample}(a) the trace \code{<ABC>} describes the model's behavioral semantics exhaustively. Obviously this single trace does not represent the semantics of \ref{fig:contExample}(c) appropriately, because of several decision points and loops. A simple formula to calculate the minimum number is shown in Eq.~\ref{eq:N}. This formula considers the size of the set of tasks ($|T|$) and is further based on the knowledge that the length of the longest possible sequence of tasks without repetition is given by $L$. The formula also factors in arbitrary task orderings (the ith power) and shorter paths (the sum). Using this formula we do not need any information about the structure of the process model.

\begin{minipage}{.4\linewidth}
	\begin{equation}
	\label{eq:N}
		N \geq \sum_{i \leq L}^{i=0} |T|^i 
	\end{equation}
\end{minipage}%
\begin{minipage}{.4\linewidth}
	\begin{equation}
	\label{eq:L}
		L \geq 2|T| 
	\end{equation}
\end{minipage}%

Assuming that the simulation engine is able to produce shorter traces, too, the formula for \code{L} is based on the idea that the longest trace without repetition (\code{|T|}) could be executed at least twice (Eq. \ref{eq:L}).\\
\indent Both formulae describe just the absolute lower bound for both dimensions. We therefore suggested to choose both parameters significantly higher. However, since $N$ increases exponentially with \code{L}, using this formula becomes expensive very fast. In practice considerably less traces are necessary, because the formula only considers tasks in arbitrary combinations without following the rules in the process model. Hence, our evaluation has a the twofold purpose to test our approach in general and to serve as a guideline for checking the quality of the transformation for a particular configuration in practice.

In order to increase the probability that all relevant traces are contained in the log, we have to ensure that all of them are generated with equally often. However, since this cannot be configured directly in the chosen tools, we only provide a simplistic configuration, which is discussed in the next subsection.

\subsection{Simulating Imperative Process Models}
The simulation technique has to be able to produce a certain number of traces of appropriate length. In contrast, simulation tools built for measuring KPIs usually use different configurable parameters~\cite{nakatumba2008,VanderAalst2015}: \textit{(i)} the Case-Arrival Process \code{(CAP)}, \textit{(ii)} the service times \code{(st)} for tasks as well as \textit{(iii)} the probabilities for choices. Since our intent is to reuse existing techniques and technologies we have to map our desired simulation parameters from Sec.~\ref{subsec:bpSim} to the implemented parameters of the particular simulation technique. 

The \code{CAP} influences the number of traces that can be generated. In order to ensure that the desired amount of traces $N$ is generated, the inter-arrival time ($t_a$) must be set to a constant value. Finally the minimum simulation duration $d$ can be calculated according to the formula $d = \frac{N}{t_a}$. To be able to create a target model with equivalent semantics, we have to make sure that the behavioral semantics of the source model is implicitly described by the log data as accurately as possible. Given a BPMN model this means that all control-flow edges are used in the simulation step.

Another influencing factor is the task \textit{service time}, i.e. the usual duration. For our purposes these service times have to be equal and constant for all tasks. Executing two tasks \code{A} and \code{B} in parallel with $st_B > st_A$ would always produce the same sequence during simulation: \code{<...AB...>}. The subsequent Declare mining algorithm would falsely introduce a \code{chainsuccession(A,B)} instead of a correct \code{coexistence(A,B)} rule. With constant and equal values the ordering is completely random which actually is one intuition of a parallel gateway.

Probability distributions are used to simulate human decisions~\cite{VanderAalst2015} at modeled gateways, which means that the outgoing edges are chosen according to a probability that follows this distribution. The probabilities for all outgoing edges of one gateway must sum up to one and, thus, the uniform-distributed probability can be calculated according to the formula $\frac{1}{n_{O,G}}$ with $n_{O,G}$ denoting the number of outgoing edges for gateway $G$. However, determining these probabilities only locally leads to significantly lower probabilities for traces on highly branched tasks. However, since we assume a completely unstructured process when developing Form.~\ref{eq:N}, in many cases we will generate far too much traces. Thus, we suggest this as an initial solution which is proved in our evaluation.

Configuring the maximum trace length $L$ is slightly more complicated. The reason is that imperative processes are very strict in terms of valid endings of instances. This involves some kind of \textit{look ahead} mechanism which is able to check whether the currently chosen step for a trace does still allow for finishing the whole process validly. Our approach restricts the trace length in an post-processing step based on a simulation of arbitrary length which is only restricted by the simulation duration. Afterwards we select only those traces which do not exceed the configured maximum trace length.

\subsection{Simulating Declarative Process Models}


The main difference between imperative and declarative process modeling languages is that the former means modeling allowed paths through the process explicitly utilizing directed-graph representations while the latter means modeling them implicitly based on rules. In \cite{Ciccio2015} the authors presented an approach for simulating Declare models based on a six-step transformation technique. First, each activity name is mapped to one alphabetic character. Afterwards, the Declare model is transformed into a set of regular expressions. For each regular expression there exists an equivalent \textit{Finite State Automaton (FSA)} which is derived in the third step. Each regular expression and, therefore, each FSA corresponds to one Declare constraint. To make sure that the produced traces respect all constraints the product of all automatons is calculated in step four. During the next step, the traces are generated by choosing a random path along the FSA product and by concatenating the characters for all passed transitions. In the sixth and last step the characters are mapped to the original activity names and the traces are written to a log file. Similar to the simulation of imperative process models, it is necessary to configure the parameters $N$ and $L$. In \cite{Ciccio2015} both parameters can be configured directly. In contrast, we have no influence on the probability distribution for the traces since the algorithm internally assumes a uniform distribution that assigns equal probabilities to all outgoing edges of each state in the FSA. Hence, again, there is a mismatch regarding the probability for highly branched paths as in the simulation for imperative models.

\subsection{Mining Imperative BPMN Process Models}
In order to complete our tool chain we have to choose a mining technique and an appropriate configuration. We selected the Flexible Heuristics Miner (FHM)~\cite{weijters2011flexible}. Though this mining algorithm first produces a so called \textit{Causal Net} that must be later converted to BPMN, the advantages outweigh the disadvantages: \textit{(i)} The algorithm is able to overcome the drawbacks of simpler approaches, such as the Alpha algorithm, \textit{(ii)} it is specialized for dealing with complex constructs. This is very important since a \hyphenquote{USenglish}{good} Declare model, which means the tasks and not too many constraints, would usually lead to a comparatively complex BPMN model. \textit{(iii)} Furthermore the algorithm is able to handle low-structured domains (LSDs), which is important since the source model is specified in Declare - a language designed \textit{especially} for modeling LSDs. 
\begin{table}[t]
\centering
\setlength\tabcolsep{3pt}
\begin{minipage}[t]{0.45\textwidth}
\centering
\begin{tabular}[t]{@{}ll@{}}
\toprule
\textbf{Param} 									& \textbf{Value}	\\ 
\midrule
\textbf{Relative-to-best} 						& 0.0      			\\
\textbf{Dependency} 							& 49.0     			\\
\textbf{Length-one/two loop} 					& 0.0   			\\
\textbf{Long distance} 							& 100.0   			\\
\textbf{All tasks connected}					& true				\\
\textbf{Long distance dep.} 					& true				\\
\textbf{Ignore loop dep. thresh.} 				& false 				  
\end{tabular}
\end{minipage}%
\qquad
\begin{minipage}[t]{0.45\textwidth}
\centering
\begin{tabular}[t]{@{}ll@{}}
\toprule
\textbf{Param} 									& \textbf{Value}	\\ 
\midrule
\textbf{Ignore Event Types}						& false      		\\
\textbf{Minimum Support} 						& 100.0     		\\
\textbf{Alpha} 									& 0.0   					
\end{tabular}
\end{minipage}
\caption{Miner configurations: FHM (l), DMM (r)}
\label{tab:minerConfig}
\vspace{-20px}
\end{table}

After choosing an appropriate algorithm a robust and domain-driven configuration is needed. A suggestion is shown in Tab.~\ref{tab:minerConfig} (left). The \textit{Dependency} parameter should be set to a value $< 50.0$ because the simulation step produces noise-free logs. It is therefore valid to assume that, according to this configuration, a path only observed once was also allowed in the source model and is therefore not negligible. The dependency value for such a single occurrence is $50$. Consequently, there is no need for setting a \textit{Relative-to-best} threshold higher than zero. If a dependency already has been accepted and the difference between the corresponding dependency value and a different dependency is lower than this threshold, this second dependency is also accepted. \textit{All tasks connected} means that all  non-initial tasks must have a predecessor and all non-final tasks must have a successor. The \textit{Long distance dependencies} threshold is an additional threshold for identifying pairs of immediately or distant consecutive tasks. Setting this value to 100.0 means, at the example of tasks \code{A} and \code{B}, that \code{A} and \code{B} must be always consecutive and must have equal frequencies. The FHM requires some special attention for short loops like \code{<...AA...>} or \code{<...ABA...>}. Setting both parameters to $0$ means that if a task has been repeated at least once in one trace, we want to cover this behavior in the target model. Consequently we have set \textit{Ignore loop dependency thresholds} to false. This configuration completes our tool chain for translating a Declare model to a trace-equivalent BPMN model.
\subsection{Mining Declarative Process Models}
Choosing an appropriate mining technique for discovering Declare models is much easier. The reason is that there are only two major approaches, one of whom is called \textit{MINERful}~\cite{CiccioM15} and the second, which is more a compilation of a mining technique and several pre- and post-processing steps, is called \textit{Declare Maps Miner} (DMM)~\cite{Maggi2012,maggi2013declarative}. We selected the latter bundle of techniques, where the decision this time is driven by a slight difference regarding quality tests~\cite{CiccioM15} and our own experiences pertaining the tool integration. Though both approaches a comparable in terms of the result quality MINERful is a bit more sensitive to the configuration of two leading parameters, namely \textit{confidence} and \textit{interest factor}. However, MINERful outperforms the DMM in terms of computation time. But according to the experiences of the authors in \cite{CiccioM15}, the latter is more appropriate in case of offline execution and is therefore also more appropriate for a highly automated tool chain. Finally the question of a target-aimed configuration is answered in Tab.~\ref{tab:minerConfig} (right). Setting \textit{Ignoring Event Types} to \textit{false} is necessary since our source model is a BPMN model and therefore may allow for parallel execution of activities. A log is based on the linear time dimension, which means that we have to distinguish between the \textit{start} and the \textit{completion} of an activity, in order to represent a parallel execution. Since Declare does not allow for parallelism explicitly, we have to interpolate this behavior through considering the event types. Of course, this leads to a duplication of the tasks, compared to the original model. The threshold for the \textit{Minimum Support} can be set to 100.0 because the log does not contain noise. The last parameter, called \textit{Alpha} avoids that some considered rules are trivially true. This can be the case, for instance, with the \code{chainprecedence(A,B)} rule in Fig.~\ref{fig:contExample} (d). If \code{B} is never executed, this rule is never violated and, therefore, trivially true. This would falsely consolidate this rule.

\FloatBarrier
\section{Evaluation} \label{sec:evaluation}
Within this section, we evaluate our approach in two stages. We do not provide an integrated implementation for our approach but describe a chain of suitable and well-established tools which are equipped with the desired functionalities.

\subsection{Implementation}
Many BPMN modeling tools provide simulation features, however, not all of them allow for the export of simulated traces. IYOPRO \cite{Uhlmann2015341} allows for importing existing BPMN models. In order to run the simulation appropriately it is possible to influence the following basic parameters: \textit{(i)} Inter-arrival times for \textit{Start Events}, \textit{(ii)} the duration of activities and \textit{(iii)} probability distributions for the simulation of decisions at gateways. Additionally it is possible to modify the overall simulated execution time. These parameters influence the number and shape of generated traces and, therefore, it is not possible to determine this parameter explicitly. In order to model the preferred trace length we have to run multiple simulations with different probability distributions for gateways. Using probability distributions already indicates that the passed paths through the process are computed randomly.

In contrast to BPMN there is only one simulation tool for Declare~\cite{Ciccio2015}. Since its primary application was the quality measurement of declarative process mining tools it is possible to specify the number of traces to generate as well as the maximum trace length explicitly. The Declare models are transformed into \textit{Finite State Automata} and paths along them are chosen randomly. We export the traces in the XES standard format. For mining processes we use the well-known \textit{ProM 6} toolkit~\cite{Dongen2005}. For BPMN it provides the \textit{BPMN Miner} extension~\cite{conforti2016bpmn}, that contains, for instance, the FHM algorithm and for Declare we use the \textit{DMM} plugin~\cite{Maggi2011}. In both cases we first import the produced artificial event logs and afterwards choose and parameterize the mining algorithm. Additionally, we use ProM's conformance checking features for analyzing the transformation quality.
\subsection{Used Evaluation Metrics}
Since the final result is generated by process mining techniques we can reuse the corresponding well-known evaluation metrics. For reasons of comprehensibility we first give a small, fuzzy introduction to these metrics~\cite{VanderAalst2011}:
\begin{enumerate}[label=({\arabic*})]
	\item \textit{Fitness} (Recall): Which proportion of the logged traces can be parsed by the discovered model?
	\item \textit{Appropriateness}(Precision): What proportion of additional behavior is allowed by a model but is not present in the log?
	\item \textit{Generalization}: Discovered models should be able to parse unseen logs, too.
	\item \textit{Simplicity}: Discovered models should be as simple as possible. An important criterion is, for instance, the model size (number of nodes/edges). 
\end{enumerate}
For the evaluation we consider only the fitness and appropriateness. The generalization of the approach as well as simplicity of a model completely depends on the used process mining algorithm and cannot be controlled by the available simulation parameters. Furthermore, measuring these dimensions independently from the source model does not give any clue whether the model complexity is caused by inappropriate mining configuration or by the complexity of the source model. Since there are no comparable approaches so far, this paper focuses on checking the principal capability of this translation system in terms of correctness - which can be measured through the two metrics for fitness and appropriateness. For our calculations in the following subsection we use the formulae for fitness and appropriateness provided in \cite{van2012replaying}.  
\subsection{Transformation Result Quality: Simple Models}\label{sec:qualiSimple}
In order to start measuring the transformation quality we first apply the introduced metrics to our small continuous examples shown in Fig.~\ref{fig:contExample}. The corresponding simulation configurations and measurement results are shown in Tab.~\ref{tab:qualiSimple}. All measurements have been produced using the corresponding ProM replay plugins with anew generated 10000 sample traces for each of the four models. The experiments have been repeated ten times and the results have been averaged.
\begin{table}[t]
\centering
\begin{tabular}{|ll|llll|llll|}
\hline
\multicolumn{1}{|c}{\multirow{2}{*}{N}} & \multicolumn{1}{c|}{\multirow{2}{*}{L}} & \multicolumn{4}{c|}{Fitness}                                                                           & \multicolumn{4}{c|}{Appropriateness}         \\ \cline{3-10} 
\multicolumn{1}{|c}{}                   & \multicolumn{1}{c|}{}                   & \multicolumn{1}{c}{(a)} & \multicolumn{1}{c}{(b)} & \multicolumn{1}{c}{(c)} & \multicolumn{1}{c|}{(d)} & \multicolumn{1}{c}{(a)} & (b) & (c)    & (d) \\ \hline
10                                      & 3                                       & 1.0                     & 1.0                     & 0.7110                  & 0.4932                   & 1.0                     & 1.0 & 0.9917 & 1.0 \\
100                                     & 3                                       & 1.0                     & 1.0                     & 0.8911                  & 0.6295                   & 1.0                     & 1.0 & 1.0    & 1.0 \\
1000                                    & 3                                       & 1.0                     & 1.0                     & 1.0                     & 0.7286                   & 1.0                     & 1.0 & 1.0    & 1.0 \\
10000                                   & 3                                       & 1.0                     & 1.0                     & 1.0                     & 0.7286                   & 1.0                     & 1.0 & 1.0    & 1.0 \\
10                                      & 6                                       & 1.0                     & 1.0                     & 0.713                   & 0.6111                   & 1.0                     & 1.0 & 0.9929 & 1.0 \\
100                                     & 6                                       & 1.0                     & 1.0                     & 0.9874                  & 0.7257                   & 1.0                     & 1.0 & 1.0    & 1.0 \\
1000                                    & 6                                       & 1.0                     & 1.0                     & 1.0                     & 0.9975                   & 1.0                     & 1.0 & 1.0    & 1.0 \\
10000                                   & 6                                       & 1.0                     & 1.0                     & 1.0                     & 1.0                      & 1.0                     & 1.0 & 1.0    & 1.0 \\
10                                      & 9                                       & 1.0                     & 1.0                     & 0.713                   & 0.6420                   & 1.0                     & 1.0 & 0.9929 & 1.0 \\
100                                     & 9                                       & 1.0                     & 1.0                     & 1.0                     & 0.7844                   & 1.0                     & 1.0 & 1.0    & 1.0 \\
1000                                    & 9                                       & 1.0                     & 1.0                     & 1.0                     & 1.0                      & 1.0                     & 1.0 & 1.0    & 1.0 \\
10000                                   & 9                                       & 1.0                     & 1.0                     & 1.0                     & 1.0                      & 1.0                     & 1.0 & 1.0    & 1.0 \\ \hline
\end{tabular}
\caption{Quality: Models (a)-(d) shown in Fig.~\ref{fig:contExample}}
\label{tab:qualiSimple}
\vspace{-20px}
\end{table}

Though the used source model for this first evaluation are very simplistic, it is possible to discern four important facts. First, the two simplest models (cf. Fig.~\ref{fig:contExample}(a) and (b)) can be transformed correctly, as expected, with a very low amount of traces of short length. Secondly, the appropriateness is almost always 100\%. The reason is that, the less traces are passed to the relevant process miner, the more restrictive is the resulting model. Both miners treat the traces as the only allowed behavior and, therefore, produce models that are as strict as the traces themselves. The third insight is that in the case of the more complex models (cf. Fig~\ref{fig:contExample}(c) and (d)) the fitness decreases. This means that for translating from BPMN to Declare more traces are required to raise the fitness, which is expected due to more execution alternatives. Finally, we have to point out that we are able to achieve 100\% fitness and appropriateness because our simulation components generate noise-free logs.
\FloatBarrier
\subsection{Transformation Result Quality: Complex Models}
\label{sec:qualiComplex}
Our second evaluation state is based on two more complex models than used in the previous subsection. The Declare source model is a model mined from real-life log data which was provided in the context of the \textit{BPI Challenge 2014}\footnote{Log available at: http://www.win.tue.nl/bpi/doku.php?id=2014:challenge}. Furthermore, the mined model has been used in \cite{Ciccio2015} as evaluation data, too. The logs have been produced in the context of customer-service-desk interactions regarding disruptions of ICT services. Consequently, the model has been mined with the Declare Maps Miner ProM extension, too. The resulting model consists of 12 tasks and 15 constraints. Our more complex BPMN model has already been used in \cite{RodriguesARBL15} in order to prove the understandability of BPMN models supported by human-readable textual work instructions. The model consists of 15 tasks and allows for 48 different paths through a bread-delivery process.
\begin{table}[t]
\centering
\setlength\tabcolsep{3pt}
\begin{minipage}[t]{0.45\textwidth}
\centering
\begin{tabular}[t]{|ll|ll|}
\hline
\multicolumn{1}{|c}{N} & \multicolumn{1}{c|}{L} & \multicolumn{1}{c}{Fitness} & \multicolumn{1}{c|}{App.} \\ \hline
100                    & 24                     & 0.6371                      & 1.0                                  \\
1000                   & 24                     & 0.8181                      & 1.0                                  \\
10000                  & 24                     & 0.9992                      & 1.0                                  \\
100000                 & 24                     & 1.0                         & 1.0                                  \\
100                    & 36                     & 0.6554                      & 1.0                                  \\
1000                   & 36                     & 0,8988                      & 1.0                                  \\
10000                  & 36                     & 0.9998                      & 1.0                                  \\
100000                 & 36                     & 1.0                         & 1.0                                  \\ \hline
\end{tabular}
\end{minipage}%
\qquad
\begin{minipage}[t]{0.45\textwidth}
\centering
\begin{tabular}[t]{|ll|ll|}
\hline
\multicolumn{1}{|c}{N} & \multicolumn{1}{c|}{L} & \multicolumn{1}{c}{Fitness} & \multicolumn{1}{c|}{App.} \\ \hline
10                     & 15                     & 0.5                         & 1.0                                  \\
100                    & 15                     & 0.6253                      & 1.0                                  \\
1000                   & 15                     & 0.7462                      & 1.0                                  \\
10000                  & 15                     & 0.8784                      & 1.0                                  \\
100000                 & 36                     & 0.9335                      & 1.0                                  \\ \hline
\end{tabular}
\end{minipage}
\vspace{5pt}
\caption{Model translation quality BPI Ch. 2014 (l), Bread deliv. process (r)}
\label{tab:evalComplex}
\end{table}
Again, we evaluated the translation quality with ten log files containing 10000 traces, respectively. The quality measurements, shown in Tab.~\ref{tab:evalComplex}, are averaged, too.

Both tables show that we ware able to translate the models to a very high degree and confirm the findings of the previous evaluation step, which means that the quality is only a matter of fitness and, thus, target models produced with too few traces tend to be overfitted, which is an expected and well-known issue in machine learning. For these two example a significant higher amount of traces is required.

Additionally, we analyzed the performance of our approach only slightly, since it is based upon techniques that have already been analyzed regarding the computation time. Our evaluation has been performed on the following hardware: Dell Latitude E6430, intel Core i7 3720QM (4 x 2.6 GHz), 16 GB DDR3 internal memory, SSD, Windows 8 (64 bit). Translating models like our small continuous examples in Fig.~\ref{fig:contExample} require only few traces and, therefore can be performed in an average time of one second. Translating our two more complex models require far more traces which leads to an average computation time of 8 (BPI Ch.) or 10 (Bread del.) seconds. For more precise and broader performance analysis, please consider the corresponding literature for the four used components~\cite{Ciccio2015,Uhlmann2015341,weijters2011flexible,maggi2013declarative}.

\section{Related Work} \label{sec:relatedwork}
This paper relates to different streams of research: process modelling approaches and process model transformation. In general, the differences between declarative and imperative process modeling approaches have been discussed in \cite{Pichler2012} where both imperative and declarative languages are investigated with respect to process model understanding. The most relevant work in the context of process model transformation between these different paradigms is \cite{Prescher2014}. The paper describes an approach to derive imperative process models from declarative models. It utilizes a sequence of steps, leading from declarative constraints to regular expressions, then to a finite state automaton and finally to a petri net. To the best of our knowledge there is no other specific approach for the translation of declarative process models. There are, however, different approaches that translate process models from one imperative language to another one, e.g., BPMN to BPEL \cite{recker2006translation}. Furthermore our work is related to the approach presented in \cite{Wimmer2007}. There the main issue of writing cumbersome model transformation rules is solved providing a transformation approach that works on examples. Similarly our approach works on exemplary models but, in contrast, these are composed using simulation techniques which means that we prevent the user from any overhead. For our transformation approach we make use of different process model simulation \cite{Ciccio2015} and process mining techniques \cite{weijters2011flexible,maggi2013declarative} that have already been mentioned and described throughout the paper.

\section{Conclusion and Future Work} \label{sec:conclusion}
The process model translation approach presented in this paper provides an alternative to classical model-to-model-translation based systems. It is based on the assumption that a representative event log can be used as a transfer medium in order to reduce the complexity of process model translation systems. In order to ensure suitability in practice we evaluated our approach on real-life data. Our evaluation showed that with a certain amount of simulated traces it is possible to cover the behavioral semantics of our exemplary source model in the log. However, in order to use the approach in real-life applications the general log expressiveness must be investigated in advance. The same applies to the expressiveness of the relevant pairs of process modeling languages. Furthermore there are pairs of languages where both provide a certain support for modeling relations beyond plain control-flow dependencies. In order to generate all relevant traces from large and highly branched BPMN models, it is necessary to find a more suitable algorithm to define appropriate probability distributions for decisions at gateways. This could be achieved, for instance, if the algorithm considers not only the number of outgoing edges of each gateway but also the branching factor of all subsequent gateways. These two improvements lead to a more accurate calculation of the maximum number of unique traces and, hence, of the number of required simulated traces.

\bibliographystyle{ieeetr}
\bibliography{literature}
\end{document}